\documentclass[conference, a4paper]{IEEEtran}
\usepackage{cite}
\usepackage{amsmath,amssymb,amsfonts}
\usepackage{graphicx}
\usepackage{textcomp}
\usepackage{tikz}
\usepackage{adjustbox}
\usepackage{multirow}
\usepackage{xcolor}
\usepackage{pgfplots}
\usepackage{subcaption}
\usepackage{booktabs}
\usepackage{algpseudocode}
\usepackage[ruled, lined, longend, linesnumbered]{algorithm2e}
\usepackage[utf8]{inputenc}
\usepackage{tikz}

\usepackage{amssymb}
\usepackage{pifont}
\begin{document}
\title{Federated Learning for Zero-Day Attack Detection in 5G and Beyond V2X Networks}

\author{\IEEEauthorblockN{1\textsuperscript{st} Abdelaziz {Amara korba}}
\IEEEauthorblockA{\textit{L3I, University of La Rochelle, France } \\
\textit{LRS, Badji Mokhtar Annaba University, Algeria}\\
}
\and
\IEEEauthorblockN{2\textsuperscript{nd} Abdelwahab Boualouache}
\IEEEauthorblockA{\textit{FSTM, University of Luxembourg} \\
Luxembourg\\
}
\and
\IEEEauthorblockN{3\textsuperscript{rd} Bouziane Brik}
\IEEEauthorblockA{\textit{DRIVE, University of Bourgogne} \\
\textit{Franche Comte, France}\\
}
\and
\IEEEauthorblockN{4\textsuperscript{rd} Rabah Rahal}
\IEEEauthorblockA{\textit{LRS, Badji Mokhtar Annaba University, Algeria} \\
}
\and
\IEEEauthorblockN{5\textsuperscript{rd} Yacine Ghamri-Doudane}
\IEEEauthorblockA{\textit{L3I, University of La Rochelle,France} \\
}
\and 
\IEEEauthorblockN{6\textsuperscript{rd} Sidi Mohammed Senouci}
\IEEEauthorblockA{\textit{DRIVE, University of Bourgogne} \\
\textit{Franche Comte, France}\\
}
}

\maketitle
\begin{abstract}
Deploying Connected and Automated Vehicles (CAVs) on top of 5G and Beyond networks (5GB) makes them vulnerable to increasing vectors of security and privacy attacks. In this context, a wide range of advanced machine/deep learning-based solutions have been designed to accurately detect security attacks. Specifically, supervised learning techniques have been widely applied to train attack detection models. However, the main limitation of such solutions is their inability to detect attacks different from those seen during the training phase, or new attacks, also called zero-day attacks. Moreover, training the detection model requires significant data collection and labeling, which increases the communication overhead, and raises privacy concerns. 
To address the aforementioned limits, we propose in this paper a novel detection mechanism that leverages the ability of the deep auto-encoder method to detect attacks relying only on the benign network traffic pattern. Using federated learning, the proposed intrusion detection system can be trained with large and diverse benign network traffic, while preserving the CAVs’ privacy, and minimizing the communication overhead. The in-depth experiment on a recent network traffic dataset shows that the proposed system achieved a high detection rate while minimizing the false positive rate, and the detection delay.
\end{abstract}


\section{Introduction}
Fifth generation (5G) and beyond (5GB) networks promise to revolutionize the transportation industry by enabling ultra-reliability with ultra-low latency and high bandwidth communications~\cite{ahmad2019security}. These advances will significantly empower many verticals, such as smart agriculture, health, and Intelligent Transportation Systems (ITS). As part of ITS, Connected and Automated Vehicles (CAVs) have been taken significant and careful considerations in 3GPP 5G standards ~\cite{9345798}. Specifically, integrating V2X communications into the 5G ecosystem has enabled innovative use cases and applications, such as advanced driving, vulnerable road user protection, and vehicle platooning~\cite{3gpp1}. Yet this progress is expected to be extended with 5GB, contributing thus to reducing traffic accidents and dramatically saving road users' lives. However, CAVs at all automation levels will face a massive vector of cyberattacks coming from 5GB technologies and leading to hazardous situations for road users. For example, Distributed and Denial of Service (DDoS) attacks have already been demonstrated to break 5G services~\cite{kuadey2021deepsecure}. But the impact of these attacks are likely to be more expansive with the integration of CAVs. More than this, cyberattacks are working continuously to develop novel tactics for breaching and breaking such systems. Facing all these challenges,  Machine Learning (ML) appears as a key cybersecurity enabler to protect 5GB-enabled CAVs~\cite{boualouache2022survey}. Various Machine Learning (ML)/ Deep Learning (DL) based Intrusion Detection Systems (ML/DL-based IDSs) have been proposed to protect vehicular networks against attacks. Most of them rely on supervised and centralized learning \cite{antibot}. Centrally training the detection model requires significant data collection and labeling, which increases the communication overhead, and may raise privacy concerns. To mitigate centralized learning limitations, collaborative ML \cite{FL} has been used, enabling thus continuous accuracy evolution and flexibility. Nevertheless, several limitations exist in early collaborative ML-based IDSs~\cite{negi2020, shu2020collaborative,zhang2018distributed}. Specifically, they generate a significant communication overhead during ML model updates and may violate data privacy, since learning nodes might share private information. To cope with the aforementioned issues, recent research ~\cite{uprety2021privacy,boualouache2022federated,liu2021blockchain,hbaieb2022federated} leveraged the potential of federated leaning (FL) paradigm, which has shown promising results in many applications. FL is a distributed ML paradigm allowing several nodes to train a global model cooperatively without sharing their datasets, avoiding thus overhead and mitigating privacy risks ~\cite{FL}. Interestingly, all existing FL-based IDSs for 5GB-enabled CAVs rely on supervised learning techniques. One important limitation of using such techniques is their inability to detect attacks different from those seen during the training phase (unseen attacks), and zero-day attacks. Another challenging issue is data imbalance, i.e., the numbers of benign and malicious traffic samples are not in the same range. Benign network traffic samples are easily available. On the other hand, malicious samples are scarce or unavailable. The lack of a thorough dataset of attack samples limits the usage of supervised techniques. Finally, most existing detection approaches assume that FL clients maintain labeled datasets that may use at each round. This assumption may not be realistic, as CAVs cannot label the network flow on every turn. 

As a CAV runs a set of well-known applications (safety, convenience, commercial, etc), their communication pattern should present a high degree of regularity so long as they are not under attack or faulty. Similarly, an attack must alter its communication pattern. Therefore, we believe it is possible to use anomaly detection techniques to model the CAV’s benign (or expected) communication pattern and detect attacks as anomalous occurrences. To overcome the aforementioned limitations of the existing IDSs, in this paper, we propose an unsupervised federated learning based IDS that leverages a deep auto-encoder model to train the detection model, relying only on benign network traffic. Thanks to federated learning, the proposed IDS can be trained with large and diverse benign network traffic, while preserving the CAVs’ privacy. The proposed IDS aggregates the detection model updates within the Multi-access Edge Computing (MEC) server to enhance the learning efficiency and minimize latency. The in-depth experiment on a recent network traffic dataset shows that the proposed system achieves a high detection rate while
minimizing the false positive rate, and the detection delay.

The remainder of this paper is organized as follows. Section~\ref{RT} describes related work. The design of our scheme is presented in Section~\ref{SOL}. Section~\ref{SIM} depicts the performance evaluation results, and finally, Section~\ref{CON} concludes the paper.

\section{Related Work} \label{RT}
Several distributed ML-based IDSs have been proposed for detecting attacks in 5GB-enabled CAVs. Negi et al.~\cite{negi2020} presented a DL-based IDS to detect anomalies in ITS based on Long Short-Term Memory (LSTM). In this work, time series data are collected by CAVs and sent to the cloud to enable the training and retraining of a global model using a cluster of servers instead of one server. Shu et al.~\cite{shu2020collaborative} proposed a collaborative IDS based on supervised DL, Generative Adversarial Networks (GANs), and Software Defined Networking (SDN). The proposed system enables distributed SDN controllers managing sub-networks of CAVs to train a global model for the whole network without directly exchanging their sub-network flows. However, both \cite{negi2020} and \cite{shu2020collaborative} raise privacy issues since datasets are shared between learning nodes. Zhang et al.~\cite{zhang2018distributed} proposed a distributed ML-based IDS that enables CAVs to directly communicate to train a global model based on supervised learning without sharing their datasets. However, peer-to-peer distributed learning generates a large overhead, degrading communication performance. Uprety et al.~\cite{uprety2021privacy} proposed a FL-based privacy-preserving collaborative IDS for CAVs. This work enables CAVs (FL clients) to train DL models on a locally labeled dataset and share their parameters with the central FL server to build a global model. Boualouache et al.~\cite{boualouache2022federated} proposed a FL-based privacy-preserving collaborative IDS based on supervised learning that leverages a set of FL servers to train the global model. Liu et al.~\cite{liu2021blockchain} proposed FL for collaborative IDS. This work suggests offloading the training to distributed vehicular edge nodes. Specifically, CAVs act as FL clients for building models based on their locally labeled datasets and Roadside Units (RSUs) for aggregating global models. Hbaieb et al.~\cite{hbaieb2022federated}  proposed SDN-FL-based IDS for CAVs. In this work, SDN controllers train local models based on labeled datasets built using data collected from CAVs, while the aggregation of global models is performed on the cloud.

Overall, existing FL-based IDSs for 5GB-enabled  CAVs~\cite{uprety2021privacy,boualouache2022federated,liu2021blockchain,hbaieb2022federated} have specifically two main limitations: (i) they are based on supervised learning which limits their effectiveness against unseen and zero-day attacks, and (ii) they assume that FL clients have labeled datasets, which in practice might be unrealistic. Considering these gaps, we propose a novel network-based IDS trained using a deep auto-encoder model. Relying only on benign network traffic, our system can detect unseen or zero-day attacks so long as they alter the benign communication pattern of the CAV. Additionally, our solution does not compromise the CAV’s privacy since it is built through federated learning.




\section{Proposed Solution}\label{SOL}
\begin{figure*}[ht!]
    \centering
    \includegraphics[scale=0.36]{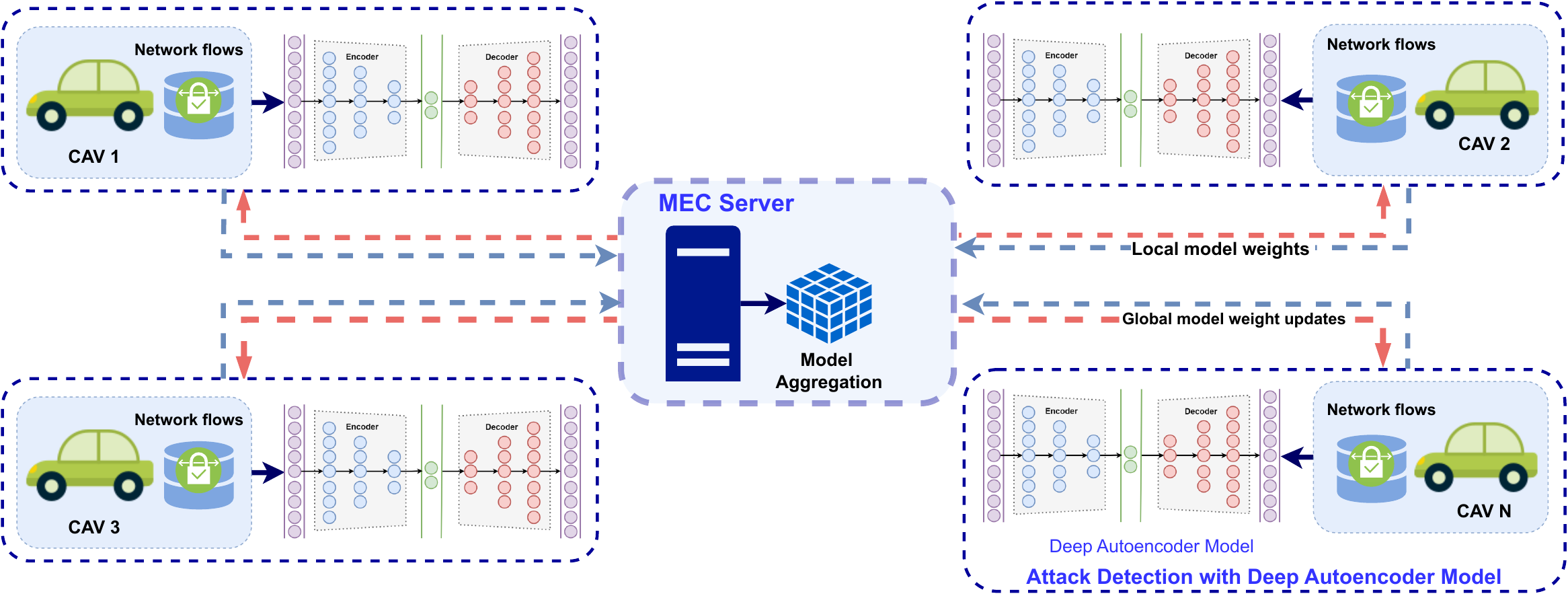}
    \caption{MEC-Enabled Federated learning architecture}
    \label{fig:archi3}
\end{figure*}

Collaborative learning allows training the model with a large amount of network traffic from diverse CAVs, while preserving data
privacy, and minimizing the communication overhead. The proposed MEC-enabled learning scheme trains the deep Auto-Encoder (AE) model in a federated way, as illustrated in Figure \ref{fig:archi3}. First, the raw captured packets are converted to flows. Then, for each flow, a set of pertinent features are calculated. Next, the local dataset of flows is fed to the AE model initially distributed by the MEC server. The training rounds are orchestrated by the MEC server and executed by the AE on the participating CAV's local dataset. 

\subsection{Flow extraction \& features engineering}
First, to identify a traffic flow, we use a combination of five properties from the packet header, including the network and the transport layer headers of the TCP/IP protocol stack. These are as follows: source IP address,
destination IP address, source port number, destination port number, and protocol. For each flow extracted, a set of features are calculated according to a given time window (ex. 100 seconds). Flow features include mainly packet header characteristics and statistics computed from the aggregating network and transport layers header information of the packets in a flow. The network features are related to: time, packets, bytes, and flag groups. The list of features as well as their description are given in table \ref{tab:tab-FlowStat}.

\begin{table*}[]
\centering
\caption{List of network features}
\label{tab:tab-FlowStat}
\begin{adjustbox}{width=0.8\textwidth}
\begin{tabular}{@{}ll@{}}
\toprule
\textbf{Feature names}                               & \textbf{Description}                                                            \\ \midrule
numHdrs                                              & Number of headers (depth) in hdrDesc                                            \\
l4Proto                                              & Layer 4 protocol (TCP, UDP, HOPOPT,…)                                           \\
dstPortClass                                         & Port based classification of the destination port name (HTTPS, Telnet, NTP, ..) \\
numPktsSnt, numByteSnt                               & Number of transmitted packets/ byte per second                                  \\
minPktSz, maxPktSz, avePktSize, stdPktSize           & Minimum/Maximum/Average/ Standard deviation of layer 3 packet size              \\
pktAsm, bytAsm                                       & Packet/ Byte stream asymmetry                                                   \\
duration                                             & Duration of the flow in seconds                                                  \\
minIAT, maxIAT, aveIAT, stdIAT                       & Minimum/ Maximum/ Average/ Standard deviation of inter-arrival time             \\
pktps/ bytps                                         & Number of sent packets/ byte per second                                         \\
ipTTLChg, ipTOS, ipFlags                             & IP TTL change count, IP Type of Service, IP aggregated flags                    \\
tcpISeqN                                             & TCP Initial Sequence Number                                                     \\
tcpSeqFaultCnt                                       & TCP sequence number fault count                                                 \\
tcpPAckCnt                                           & TCP packet ack count                                                            \\
tcpFlwLssAckRcvdBytes                                & TCP flawless ack received bytes                                                 \\
tcpInitWinSz, tcpAveWinSz, tcpMinWinSz, tcpMaxWinSz, & TCP initial/ average / minimum/ Maximum effective window size                   \\
tcpWinSzDwnCnt, tcpWinSzUpCnt                        & TCP effective window size change down/up count                                  \\
tcpWinSzThRt                                         & TCP packet count ratio below window size WINMIN threshold                       \\
tcpFlags                                             & TCP aggregated protocol flags (cwr, ecn, urgent, ack, push, reset, syn, fin)    \\
tcpAnomaly                                           & TCP aggregated header anomaly flags                                             \\
tcpOptions                                           & TCP aggregated options                                                          \\
tcpMSS                                               & TCP maximum segment size                                                        \\
tcpEcI                                               & TCP estimated counter increment                                                 \\
tcpUtm, tcpBtm                                       & TCP estimated up/boot time                                                      \\
tcpSSASAATrip                                        & TCP trip time SYN, SYN-ACK Destination | SYN-ACK, ACK Source                    \\
tcpRTTAckTripMin, tcpRTTAckTripMax, tcpRTTAckTripAve & TCP ACK minimum/ maximum/ average trip time                                     \\
tcpRTTAckTripJitAve, tcpRTTAckJitAve                 & TCP ACK trip/ round trip time average jitter                                    \\
tcpRTTSseqAA                                         & TCP round trip time \{SYN, SYN-ACK, ACK\} and \{ACK-ACK\}                       \\ \bottomrule
\end{tabular}
\end{adjustbox}
\end{table*}

\subsection{Modeling benign network traffic pattern}
The Auto-Encoder (AE) model \cite{ae} is an unsupervised model that compresses input vectors as code vectors using a set of recognition weights and then converts back to $m$ $(m < d)$ number of neurons reconstructed input vectors using a set of generative weights. There are two major parts in an AE architecture: the encoder and the decoder. The encoder reduces the dimension of the input vectors ($x_{i} \in R^{d}$) to numbers of neurons that form the hidden layer. The activation of the neuron $i$ in the hidden layer is given by: 

\begin{equation}
    \label{eq1}
h_{i}=f^{_{\theta }}(x)=s(\sum_{j=1}^{n}W_{ij}^{input}x_{j}+b_{i}^{input})
\end{equation}

where $x$ is the input vector, $\theta$ is the parameters $\begin{Bmatrix}
W^{input},b^{input}
\end{Bmatrix}$, $W$  is an encoder weight matrix of dimension $m\times d$, while $b$ is a bias vector of dimension $m$. Thus, the input vector is encoded to a vector with fewer dimensions. The decoder maps the low-dimensional hidden representation $h_{i}$ to the original input space $R^{d}$  by the same transformation as the encoder. The function of mapping is as follows:
\begin{equation}
    \label{eq2}
x_{'}^{i}=g_{\theta' }(h) = s(\sum_{j=1}^{n}W_{ij}^{hidden}h_{j}+b_{i}^{hidden})
\end{equation}

The set of decoder parameters is $\theta'(W^{hidden}h_{j}+b^{hidden})$. The objective of an autoencoder is to minimize the reconstruction error relative to $\theta$ and $\theta'$ :

\begin{gather}
        \label{eq3}
\theta ^{*},\theta ^{'*}=arg_{\theta,\theta ^{'}} min \frac{1}{n}\sum_{i=1}^{n}\varepsilon (x_{i},x_{i}^{'}) \\
=arg_{\theta,\theta ^{'}} min \frac{1}{n}\sum_{i=1}^{n}\varepsilon (x_{i},g_{\theta ^{'}}(f_{\theta}(x_{i})))
\end{gather}

The reconstruction error is utilized as the anomaly score. Network flows with significant reconstruction errors are regarded as malicious flows (anomalies). Only benign flows are used to train the AE model. After training, the AE model will reconstruct benign flows exceptionally well, but not malicious flows that it has never seen. \textbf{Algorithm} \ref{algo_AE} illustrates the anomaly detection process using the reconstruction errors of the AE model. The threshold $\alpha$ is the sum of the sample's mean squared error (MSE) median and the sample's five times the MSE median absolute deviation (MAD) over the validation set. MAD uses the deviation from the median, which is less likely to be skewed by outlier values.


\begin{algorithm}[h!]

    \caption{AE-based Anomaly Detection}
    \label{algo_AE}
    \textbf{BEGIN} \\
    \textbf{\textit{PHASE 1: Flow extraction \& Preprocessing}}  \\ 
    \textbf{INPUT}: $pkts$: raw packets, $TW$: Time window (s) \\
     Extract flow from $pkts$ according to the $TW$ \\
     $Data\leftarrow$ Calculate $l$ features vectors (see table \ref{tab:tab-FlowStat} for features list) \\
     $X_{Tr},X_{V},X_{Te}\leftarrow$ Splitting $Data$

  \textbf{\textit{  \#\# PHASE 2: Training the AE Model  }} \\ 
    \textbf{INPUT}: $X_{Tr}$ : Train dataset, $X_{V}$: Validation dataset \\
    $\phi, \theta \leftarrow$ train the AE on $X_{Tr}$ \\
     \For{$i \in \{1,...,N\}$}{
         $RE_{V}[i]=\left \| x_{V}^{(i)}-g_{\theta }(f_{\phi}(x_{V}^{(i)})) \right \|$  
                } 
     $\alpha = \widetilde{MSE}_{RE_{V}} + 5\times median(RE_{V}[i] - median_{RE_{V}})$ \\

\textbf{\textit{\#\# PHASE 3: Testing} } \\ 
    \textbf{INPUT}: $X_{Te}$: Test dataset, $\alpha $: Threshold \\
    
    \For{$i \in \{1,...,N\}$}{
         $RE[i]=\left \| x_{Te}^{(i)}-g_{\theta }(f_{\phi}(x_{Te}^{(i)})) \right \|$  \\
         \eIf{$RE[i]> \alpha$}{$x_{Te}^{(i)}$ is a malicious flow}{$x_{Te}^{(i)}$ is a benign flow}
                }
  	 \textbf{END}
\end{algorithm}


\subsection{MEC-enabled federated training process}

First, the MEC server initializes the learning parameters of the shared model in terms of neural network configuration (number of layers, number of neurons, activation functions, etc.), batch size, learning rate, number of epochs, etc. It then shares such parameters with the participant CAVs. Each CAV computes local updates on top of both the AE model and its local data. Once done, each CAV sends its local model's parameters (weights) to the MEC server. The latter aggregates the received local models to generate a global learning model, before sending it back to the involved CAVs, in order to initiate a new training round. 

In our study, the federated learning problem across multiple CAVs is formulated as a federated optimization problem and resolved using the FedAvg algorithm \cite{FedAvg}. Indeed, using its local data, each CAV calculates the average gradient on top of the model $w$ for a corresponding training round $r$. Thereafter, each CAV performs a local gradient descent on the currently used model with its own data. On the other hand, the MEC server aggregates these local updates and transfers back the global model to the CAV collaborators. This process is repeated during a number of rounds, defined initially by the MEC server. \textbf{Algorithm \ref{FL}} illustrates the main steps performed by both the central MEC server and each participating CAV (Client). 
\begin{algorithm}[ht!]

    \caption{Federated Averaging Algorithm}
 
    \label{FL}

    \textbf{BEGIN} \\
    \textbf{Variables}: \textit{K}: index of clients,\textit{ B}: local batch size, \textit{E}: number of local epochs, \textit{$\eta$} : learning rate   \\

    \textbf{ClientUpdate:} \# \\ 
                $\beta \leftarrow$ (split $\mathit{P_{k}}$ into batch of size $\mathit{B}$ ) \\
                \For{each local epoch $i \in \{1,...,E\}$}{
                 \For{batch $b \in \textit{B}$}{
                $w\leftarrow \textit{w} - \eta \bigtriangledown \mathit{\mathfrak{\mathit{l}}}
                 (w;\mathit{b})$ \\
                }
                return $w$to server
                }

    \textbf{Server executes:} \\
     Initialize ${w}_{0}$ \\
            
              \For{ each round $t \in \{1,...,N\}$}{
                $m\leftarrow$ max(C. K, 1) \\
                $S_{t}\leftarrow$ (random set of m clients) \\
                \For{each client $k \in S_{t} $ in parallel}{
                   $w_{t+1}^{k}\leftarrow ClientUpdate(k,w_{t})$ \\
                    $w_{t+1}\leftarrow \sum_{k=1}^{k}\frac{n_{k}}{n} w_{t}^{k}+1$ \\
                }
              }

  	 \textbf{END}
\end{algorithm}


\section{Performance Evaluation}\label{SIM}
In this section, we first briefly describe the dataset \cite{VDoS} used in this research. Then, we present and discuss in detail the detection performances of the proposed system. Finally, we compare our approach with supervised and centralized approaches.

\subsection{Dataset preprocessing \& features enginnering}
To the best of our knowledge, VDoS \cite{VDoS} is the only publicly available dataset that includes benign and malicious traffic generated based on a realistic testbed. The network traffic was gathered in three different settings: urban, rural, and highway. The experimental environment included two vehicles, 3 physical machines, 4 virtual machines, 2 access points, a 4G modem, and two Cisco antennas. Common user applications such as Google Maps, YouTube, social networks, and other real-time applications (video/audio calls) have been run to generate benign network traffic. To generate malicious traffic, three Kali-Linux machines run three scenarios of DoS attack: UDP Flood, SYN Flood, and Slowloris packets alternately. In this research, we do not consider the third scenario, because we believe it is quite unusual that a CAV may hosts a web server. For further details about the testbed and the dataset generation please refer to \cite{VDoS}.  

\begin{table}[]
\centering
\caption{Dataset samples distribution}
\begin{tabular}{@{}llll@{}}
\toprule
& Highway & Rural & Urban \\ \midrule
Normal         & 48303   & 32303 & 65742 \\
SynFlood       & 65790   & 65790 & 67043 \\
UdpFlood       & 65790   & 65790 & 65790 \\ \bottomrule
\end{tabular}
\label{tab:distSamp}
\end{table}

To extract flows and calculate features from raw traffic (PCAP files), we developed some scripts using Tranalyzer flow traffic exporter \cite{tranalyzer}. We tried several time windows (TW) to sample the network traffic. Having achieved similar performance, and taking into account that a smaller TW allows faster detection, we fixed the TW to 1s. Table \ref{tab:distSamp} presents the number of  benign and malicious flows extracted for each type of environment. Pearson correlation filter is then used to discard highly correlated features ( $> 95\%$).

\subsection{Experimental results}
We trained and tested the proposed system in the Google Colab cloud environment. We used the Pytorch package to implement the local and federated learning models. We implemented a deep auto-encoder with three hidden layers (50\% dimension decrease from one layer to another). The binary Mean Squared Error (MSE) loss function was used. Table \ref{tab:param} illustrates the hyperparameter setup used for local and federated learning. 
To evaluate the detection performance of our solution, in addition to the false positive rate (FPR), we considered the following metrics :  

\begin{equation}
\text{Accuracy} =\frac{TP+TN}{TP+TN+FP+FN}
\label{eq:acc}
\end{equation}

\begin{equation}
\text{F1-Score} = \frac{2TP}{2TP+FP+FN}
\label{eq:fscore}
\end{equation}

\begin{table*}[]
\centering
\caption{Hyperparameters values}
\label{tab:param}
\begin{adjustbox}{width=0.7\textwidth}

\begin{tabular}{@{}lll@{}}
\toprule
\textbf{Hyperparameter} & \textbf{Value} & \textbf{Description}                                                         \\ \midrule
lr                      & 0.012          & Learning rate for the deep AE                                          \\
Nb\_clients             & 10             & Number of clients                                                            \\
Nb\_selected            & 3-6              & Number of clients we choose for train                                        \\
Batch\_size             & 128            & Defines the dataset size in each training iteration                          \\
R\_samp\_sz             & 1000           & Choose some data from the train set to retrain the data from trained model   \\
Nb\_rounds              & 20             & Total number of communication rounds for the global model to train           \\
Epochs                  & 15             & For train client model                                                       \\
Nb\_retrain\_epochs     & 5              & Total number for retrian the global server after receiving the model weights \\
Nb\_local\_epochs       & 50             & Only for the local deep AE training                                \\ \bottomrule
\end{tabular}
\end{adjustbox}
\end{table*}

\begin{figure}[ht!]
    \centering
    \includegraphics[scale=0.57]{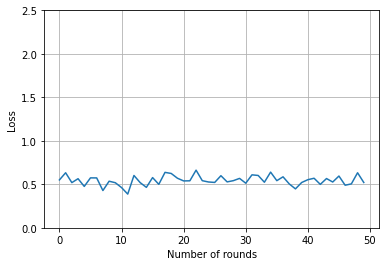}
    \caption{Loss Vs Nb. rounds}
    \label{fig:loss}
\end{figure}

Overall, the proposed solution presented a high detection rate with a low false positive rate. The detection model was able to discriminate between the reference benign traffic profile and malicious network traffic. Table \ref{tab:all_res} shows how the average accuracy, F1-score, and false positive rate may change by varying the number of participating CAVs. Using just three CAVs allowed for nearly the same detection performances while cutting the training time by about 30\%. We can see from figure \ref{fig:loss}, that there is no further improvement in terms of loss function score beyond the $11^{th}$ round. This demonstrates that the detection model did not require a large number of communication rounds to converge. 

\begin{figure}[htp]
\centering
\pgfplotsset{width=8cm,height=7.3cm,compat=1.16}
\begin{tikzpicture}
\begin{axis}[
    ybar,
    enlarge x limits=1,
    ylabel={},
    symbolic x coords={SynFlood, UdpFlood},
    xtick=data,
     enlarge y limits={upper,value=0.2},
    nodes near coords,
    enlarge y limits={rel=0.2,upper},
    ybar,
    legend style = {
     at = {(.47,.7)},
	 anchor=south,
	 column sep = 1ex
	 },
    every node near coord/.append style={rotate=90, anchor=west},
    bar width = 12pt,
    enlargelimits=0.03,
    x tick label style={align=center},
    enlarge x limits=0.4,
    ymax=110,
    bar width=12pt,
        ymin=0,
        ymajorgrids = true,
    ] 			
\addplot coordinates {(SynFlood,89.13) (UdpFlood,88.94)};

\addplot coordinates {(SynFlood,90.05) (UdpFlood,89.9)};

\addplot coordinates {(SynFlood,6.59) (UdpFlood,7.20)};

\legend{Accuracy,F1-Score,FPR}
\end{axis}
\end{tikzpicture}
\caption{Predictive performances per attack scenario }
\label{bar:type}
\end{figure}
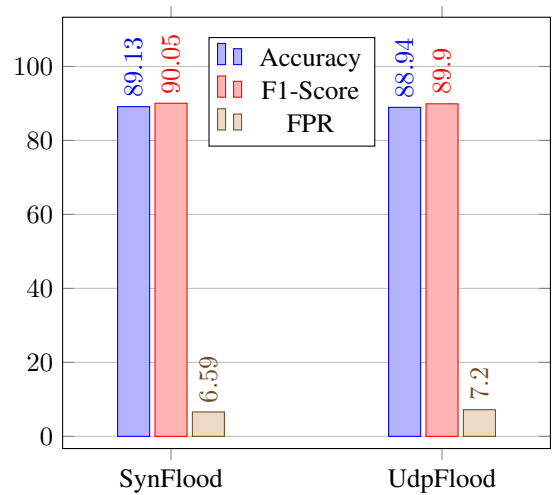

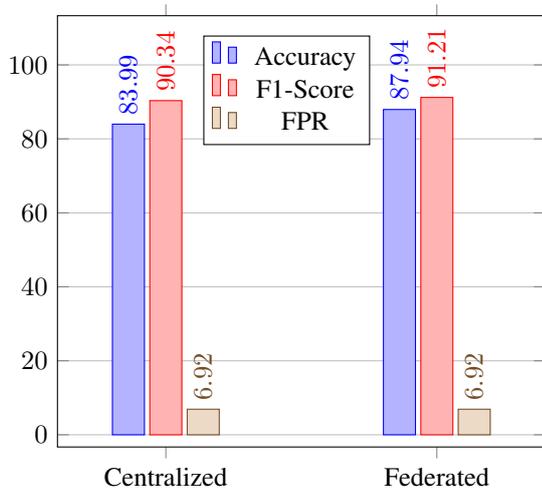
\begin{figure}[htp]
\centering
\pgfplotsset{width=8cm,height=7.3cm,compat=1.16}
\begin{tikzpicture}
\begin{axis}[
      ybar,
    enlarge x limits=1,
    ylabel={},
    symbolic x coords={Centralized, Federated},
    xtick=data,
     enlarge y limits={upper,value=0.2},
    nodes near coords,
    enlarge y limits={rel=0.2,upper},
    ybar,
    legend style = {
     at = {(.47,.7)},
	 anchor=south,
	 column sep = 1ex
	 },
    every node near coord/.append style={rotate=90, anchor=west},
    bar width = 12pt,
    enlargelimits=0.03,
    x tick label style={align=center},
    enlarge x limits=0.4,
    ymax=110,
    bar width=12pt,
        ymin=0,
        ymajorgrids = true,
    ]  				
\addplot coordinates {(Centralized,83.99) (Federated,87.94)};

\addplot coordinates {(Centralized,90.34) (Federated,91.21)};

\addplot coordinates {(Centralized,6.92) (Federated,6.92)};

\legend{Accuracy,F1-Score,FPR}
\end{axis}
\end{tikzpicture}
\caption{Performances comparison between centralized and federated learning}
\label{bar:comp}
\end{figure}

As can be observed from figure \ref{bar:type}, the proposed system performed better against the SynFlood attack. This can be explained by the relatively large number of TCP-related features (compared to UDP) included in the features vector.
Although 99.99\% F1-Score and 0.01\% FPR rate have been achieved using supervised learning, specifically, decision tree in \cite{VDoS}, our system shows comparable detection performances using only benign traffic. To get a good idea of how well the proposed model worked, we compared it to a centralized model that was trained with the whole dataset. The results comparison depicted in figure \ref{bar:comp} shows that the federated model yielded remarkably similar detection performances as the centralized model.

\section{Conclusion} \label{CON}

New attack vectors have emerged from the integration of V2X communication in the 5G ecosystem, which may lead to hazardous situations for road users. Most of the existing IDSs in 5G-V2X are either unable to detect emerging zero-day attacks because they rely on supervised learning, or do not meet privacy requirements due to data collection required for centralized learning. To tackle these limits, we proposed in this paper a new IDS based on a deep auto-encoder model, which leverages the predictability of benign network traffic to detect attacks. Relying on federated training orchestrated by the MEC server, the proposed IDS did not require any data collection or labeling. Through in-depth experiments on a recent dataset, we have shown that the proposed IDS provides high performance even with few communication rounds and a short TW sampling, which allows fast training and low detection delay. 
In future work, we plan to evaluate the proposed IDS on other non-Identical Independent Distribution (non-IID) datasets including more sophisticated and recent attacks. 


\begin{table}[]
\caption{Evaluation of detection performances}
\label{tab:all_res}
\centering
\begin{tabular}{@{}ccccc@{}}
\toprule
\textbf{Nb. Clients} & \textbf{Accuracy} & \textbf{F1-Score} & \textbf{FPR} & \textbf{Time   (mn)} \\ \midrule
10                       & 87.94\%           & 91.21\%           & 6.95\%       & 14.80                         \\
8                        & 87.94\%           & 91.22\%           & 6.98\%       & 11.53                         \\
6                        & 87.95\%           & 91.23\%           & 7.06\%       & 9.32                          \\
3                        & 87.94\%           & 91.21\%           & 6.92\%       & 4.43                          \\ \bottomrule
\end{tabular}
\end{table}


\section*{Acknowledgment}

This work was supported by the 5G-INSIGHT bilateral project (ID: 14891397) / (ANR-20-CE25-0015-16), funded by the Luxembourg National Research Fund (FNR), and by the French National Research Agency (ANR).


\bibliographystyle{unsrt}
\bibliography{ref}

\end{document}